\documentclass[aps,prd,onecolumn,groupedaddress,nofootinbib,superscriptaddress]{revtex4}
\pdfoutput=1

\usepackage{amsmath}
\usepackage{bbm}
\usepackage{float}
\usepackage{graphicx}	
\usepackage{hyperref}
\usepackage{mathtools}
\usepackage{cancel}

\usepackage[normalem]{ulem}

\usepackage{multirow}
\usepackage{rotating}
\usepackage{subcaption}

\def\lsim{\raise0.3ex\hbox{$\;<$\kern-0.75em\raise-1.1ex
\hbox{$\sim\;$}}}
\def\gsim{\raise0.3ex\hbox{$\;>$\kern-0.75em\raise-1.1ex
\hbox{$\sim\;$}}}

\begin{document} 

\hfill{FERMILAB-PUB-21-446-T, NUHEP-TH/21-10}

\title{ Measuring tau neutrino appearance probability via unitarity}

\author{Ivan Martinez-Soler}
\email{ivan.martinezsoler@northwestern.edu}
\affiliation{Theoretical Physics Department, Fermi National Accelerator Laboratory, P.O. Box 500, Batavia IL 60510, USA}
\affiliation{Department of Physics and Astronomy, Northwestern University, Evanston, IL 60208, USA}
\affiliation{Colegio de F\'isica Fundamental e Interdisciplinaria de las Am\'ericas (COFI), 254 Norzagaray street, San Juan, Puerto Rico 00901}
\author{Hisakazu Minakata}
\email{hisakazu.minakata@gmail.com}
\affiliation{Center for Neutrino Physics, Department of Physics, Virginia Tech, Blacksburg, Virginia 24061, USA}

\begin{abstract}
  We propose a {\em unitarity method} for determining $\tau$ neutrino appearance probability $P(\nu_{\mu} \rightarrow \nu_{\tau})$ in long-baseline (LBL) accelerator experiments and atmospheric neutrino observations. When simultaneous {\it in situ} measurements of $P(\nu_{\mu} \rightarrow \nu_{\mu})$ and $P(\nu_{\mu} \rightarrow \nu_{e})$ proceed, as is typical in the LBL experiments, one can use unitarity to ``measure'' $P(\nu_{\mu} \rightarrow \nu_{\tau})$. A theorists' toy analysis for the model-independent determination of $P(\nu_{\mu} \rightarrow \nu_{\mu})$ and $P(\nu_{\mu} \rightarrow \nu_{e})$ is presented by using the NOvA data. It is shown in our analysis that $\lsim$5\% (8\%) measurement of $\tau$ neutrino appearance probability in neutrino (antineutrino) mode is possible in the peak region $1.5 \lsim E_\nu \lsim 2.5$ GeV. The $\nu$SM-independent nature of determination of the probabilities is emphasized. 
\end{abstract}


\maketitle

\section{Introduction}
\label{sec:introduction}

There exists a prevailing feeling in our community that the third generation is special among the fundamental fermions. It is exemplified, in particular, by the exceedingly large mass of the top quark~\cite{Zyla:2020zbs,Defranchis:2021eos}. But, even before the top quark was discovered~\cite{CDF:1995wbb,D0:1995jca} signaling its exceptionally large mass, people examined, for example, the possibility that the Higgs field conceals its origin which comes from much higher energy scale represented by the $t \bar{t}$ condensation~\cite{Nambu:1989jt,Miransky:1989ds,Bardeen:1989ds}. In more contemporary contexts, if the Higgs sector is the most likely place as portal of new physics beyond the Standard Model (SM) of particle physics, the third-generation fermions could be the best source for such information due to their strongest couplings to the Higgs boson~\cite{CMS:2019rvj,Zyla:2020zbs}. 

Among the third-generation fermions tau neutrino may be the least studied one. From now on, our discussion anticipates understanding that the observed neutrino masses are embedded into the SM, a theory which will be dubbed as the ``$\nu$SM''. So far, only a handful $\nu_{\tau}$ events had been identified. $\tau$ neutrino has first been seen experimentally in an event-by-event basis by the DONuT Group~\cite{DONuT:2007bsg}. With the use of the $\nu_{\mu}$ beam from CERN, the $\nu_{\tau}$ appearance events have first been identified by the OPERA experiment~\cite{OPERA:2018nar}. Then both experiments looked for the ``kink'' events characteristic to a $\tau$ decay in nuclear emulsion. By using the statistically enriched $\nu_{\tau}$ samples of the atmospheric neutrinos, the charged-current (CC) $\nu_{\tau}$ cross section has been measured by Super-Kamiokande (Super-K) with 21\% uncertainty~\cite{Super-Kamiokande:2017edb}, while IceCube's Deep Core measured CC + NC (neutral current) cross section with about 50\% uncertainty~\cite{IceCube:2019dqi}. 

It should be emphasized that we are now in a very good, timely position: Soon we will have intense $\tau$ neutrino beams at the far detectors in the next-generation accelerator long-baseline (LBL) experiments, Tokai-to-Hyper-Kamiokande (T2HK)~\cite{Hyper-Kamiokande:2018ofw} and the Deep Underground Neutrino Experiment  (DUNE)~\cite{DUNE:2020ypp}. Thanks to the large mixing angle $\theta_{23} \sim \pi/4$, about half of the $\mu$ neutrino beam from J-PARC and LBNF, respectively, will be transformed into the $\tau$ neutrino beam at the far detectors, Hyper-K and DUNE. Because of availability of CC production of $\tau$ leptons due to its higher beam energy, DUNE must offer the best place for exploring $\tau$ neutrino physics. Naturally, this fact is receiving keen interests in the community, see e.g., Refs.~\cite{DeGouvea:2019kea,Machado:2020yxl,Ghoshal:2019pab} and the references cited therein. We should also mention that large samples of the atmospheric neutrinos taken in these big detectors will also do the same physics, with the likely chance of much improving the existing Super-K results mentioned above. 

To facilitate the $\tau$ neutrino physics in the far detectors DUNE and Hyper-K in their full strength we must resolve one particular problem. As indicated in Eq.~\eqref{dN-dEnu-alpha}, for example, the energy distribution of leptons produced by CC reactions depends on the two unknowns, the $\nu_{\mu} \rightarrow \nu_{\alpha}$ oscillation probability and the $\nu_{\alpha}$ nucleus CC cross sections. This statement is made under the assumption that the initial $\nu_{\mu}$ flux is known. To measure the $\nu_{\tau}$ cross sections we have to know the probability $P(\nu_{\mu} \rightarrow \nu_{\tau})$, and vice versa. 

One may argue that at the present stage one can use $P(\nu_{\mu} \rightarrow \nu_{\tau})$ calculated by the $\nu$SM three-flavor mixing scheme to measure the $\nu_{\tau}$ CC cross sections. It is a sensible attitude given the current large errors in $\nu_{\tau}$ cross sections. But, when we start searching for new physics beyond the $\nu$SM in the $\nu_{\tau}$ sector, much better accuracies would be required. In this era, we must keep in mind the possibility that it would show up both in the $\nu_{\tau}$ appearance probability and the $\nu_{\tau}$ induced CC reactions. Looking for new physics effects in the tau lepton production under the assumption of no new physics in the appearance probability $P(\nu_{\mu} \rightarrow \nu_{\tau})$ (or, vice versa) may miss the key features of the phenomenon. 

In this paper we propose the {\em ``unitarity method''} for model-independent determination of the appearance probability $P(\nu_{\mu} \rightarrow \nu_{\tau})$. The idea is very simple, use unitarity assuming $\nu$SM-independent measurements of the probabilities $P(\nu_{\mu} \rightarrow \nu_{e})$ and $P(\nu_{\mu} \rightarrow \nu_{\mu})$, see section~\ref{sec:unitarity-method}. For an existence proof, we present an analysis for a model-independent extraction of the probabilities $P(\nu_{\mu} \rightarrow \nu_{e})$ and $P(\nu_{\mu} \rightarrow \nu_{\mu})$ from the data. While it is certainly at the level of ``theorists' toy'' analysis, we hope that it triggers the experimentalists' interests in measuring $P(\nu_{\mu} \rightarrow \nu_{\tau})$, and eventually leads them to the real analysis. 
We emphasize that determination of everything by experimental measurements in a model-independent manner must be the ultimate goal of experimental physics. 

\section{Unitarity method for determining $P(\nu_{\mu} \rightarrow \nu_{\tau})$}
\label{sec:unitarity-method}

In this section we describe the unitarity method for determining the appearance probability $P(\nu_{\mu} \rightarrow \nu_{\tau})$. In most of the LBL accelerator $\nu_{\mu}$ beam experiments, including T2K~\cite{T2K:2019bcf}, NOvA~\cite{NOvA:2019cyt}, DUNE~\cite{DUNE:2020ypp}, and T2HK~\cite{Hyper-Kamiokande:2018ofw}, the experimental data are, and will be taken primarily in both the $\nu_{\mu} \rightarrow \nu_{\mu}$ and $\nu_{\mu} \rightarrow \nu_{e}$ channels simultaneously. The same statement applies to the atmospheric neutrino observation even though the event characterization, e.g., identification of initial and final neutrino flavors, would be much more involved in some cases. Then, by using unitarity 
\begin{eqnarray} 
&& 
P(\nu_{\mu} \rightarrow \nu_{e}) + 
P(\nu_{\mu} \rightarrow \nu_{\mu}) + 
P(\nu_{\mu} \rightarrow \nu_{\tau}) = 1,
\label{unitarity-relation} 
\end{eqnarray}
one can ``measure''  the $\nu_{\tau}$ appearance probability $P(\nu_{\mu} \rightarrow \nu_{\tau})$. 

It is quite possible that many people thought about this or the similar ideas related to this. In that case, the present paper may add little new. But, to the best of our knowledge, the unitarity method has never been presented explicitly in an organized way, and it prompted us to write this paper. As we will learn in our discussions below, there exist many things to be understood in this method. We hope our considerations in this paper urge experimentalists to think about the unitarity method for measuring $P(\nu_{\mu} \rightarrow \nu_{\tau})$, because the real analysis can only be done by people inside the experimental collaborations. 

\subsection{Does assuming unitarity imply assuming the $\nu$SM?} 
\label{sec:unitarity-SM} 

One might ask: Is it not true that assuming unitarity is essentially equivalent to usage of the $\nu$SM expression of the probability $P(\nu_{\mu} \rightarrow \nu_{\tau})$? The answer is {\em No, not at all}. That is, unitarity is much more robust and it should generally hold. Toward having a clear cut discussion, we must first understand unitarity on generic ground:
\begin{itemize}
\item 
If only the three active neutrinos span the complete state space of neutral leptons, unitarity must hold in a model-independent manner.
There is no way to go outside the complete state space during propagation, assuming absence of inelastic scattering, absorption, etc., and hence neutrino evolution must be unitary.\footnote{
If there exists only the three active neutrinos in our world, unitarity violation can occur at the initial and final projections of the mass eigenstates from/to the flavor states, as can be seen explicitly in Ref.~\cite{Martinez-Soler:2018lcy}. } 

\end{itemize}
\noindent
Therefore, unitarity holds even in the case that neutrinos have additional interactions such as the non-standard interactions (NSI)~\cite{Wolfenstein:1977ue,Ohlsson:2012kf,Miranda:2015dra,Farzan:2017xzy} in propagation, which are under active search by various experimental methods which produced the numerous constraints~\cite{Antusch:2008tz,Biggio:2009nt,Esteban:2018ppq}. 

\subsection{How could non-unitarity come in? }
\label{sec:sterile}

Then, the question might be: 
In what circumstances can the three-neutrino unitarity be violated? The simplest answer would be existence of the fourth, sterile neutrino, which may be indicated by LSND and MiniBooNE~\cite{LSND:2001aii,MiniBooNE:2013uba}. For an overview see e.g., Refs.~\cite{Dasgupta:2021ies,Dentler:2018sju,Conrad:2013mka}. If this and the similar two or three sterile scenarios are confirmed experimentally by the various experimental methods, e.g., described in Refs~\cite{Machado:2019oxb,JSNS2:2021hyk}, our unitarity method for $P(\nu_{\mu} \rightarrow \nu_{\tau})$ has to be revised. 

Yet, all is not lost. Typically, there are two possibilities: (1) the unitarity method is still valid under the certain conditions, and (2) the unitarity method can be amended in such a way that it is valid as in the no sterile case. To discuss the first case (1), let us define the non-unitarity parameter $\xi$
\begin{eqnarray} 
&& 
\xi (E) \equiv 
1 - \left[ P(\nu_{\mu} \rightarrow \nu_{e}) + P(\nu_{\mu} \rightarrow \nu_{\mu}) + P(\nu_{\mu} \rightarrow \nu_{\tau}) \right], 
\label{xi-def} 
\end{eqnarray}
If the error of obtained $P(\nu_{\mu} \rightarrow \nu_{\tau})$ is larger than a few times $\xi$, we can ignore the issue of non-unitarity by the sterile neutrino for the moment, because the probability leaking to the sterile sector is smaller than the reachable accuracy for $P(\nu_{\mu} \rightarrow \nu_{\tau})$. 

In the case (2) we assume that the sterile neutrino masses and the mixing parameters can be measured such that a modified unitarity relation $P(\nu_{\mu} \rightarrow \nu_{e}) + P(\nu_{\mu} \rightarrow \nu_{\mu}) + P(\nu_{\mu} \rightarrow \nu_{\tau}) + \sum_{i} P(\nu_{\mu} \rightarrow \nu_{S i}) = 1$ can be set up, where the sterile label $i$ runs over a few sterile neutrinos. To the extent we know $P(\nu_{\mu} \rightarrow \nu_{S i})$ well, the modified unitarity method should work better. The method works if the error in of $P(\nu_{\mu} \rightarrow \nu_{\tau})$ is comparable or larger than the estimated errors of $\sum_{i} P(\nu_{\mu} \rightarrow \nu_{S i})$. 

\subsection{More generic scenarios for non-unitarity}
\label{sec:non-unitarity}

More generically, if there exists an extra sector which is somehow isolated from the $\nu$SM one, but has a contact by having a weak coupling or small mixing with the three neutrino species, our three neutrino system is approximately unitary, but not perfectly. The most well known example for the extra sector is the heavy right-handed (RH) neutrinos in the seesaw model of neutrino masses~\cite{Minkowski:1977sc,Yanagida:1979as,Gell-Mann:1979vob,Glashow:1979nm,Mohapatra:1979ia}. In the three active plus three RH neutrino model, the $6 \times 6$ flavor mixing matrix is unitary, but $3 \times 3$ sub-matrix for the active neutrinos is not. But in the original scenario, since the RH neutrinos are so heavy, $m_{RH} \sim10^{15}$ GeV, the non-unitarity of the flavor mixing matrix for the three active neutrinos is practically undetectable. 

In fact, much less model-dependent descriptions for more generic unitarity violation (UV) scenarios exist for beyond $\nu$SM new physics both at high scale $\gg m_{W}$~\cite{Antusch:2006vwa} and low scale $\ll m_{W}$~\cite{Fong:2016yyh}. For the terminology of high- and low-scale UV see Ref.~\cite{Fong:2016yyh}. In high-scale UV, the UV effect has to be small because the prevailing weak $SU(2) \times U(1)$ structure allows us to use the charged lepton probe to constrain UV effect in the neutrino sector~\cite{Antusch:2006vwa}. It indeed entails the severe bounds on UV~\cite{Antusch:2014woa,Escrihuela:2015wra,Fernandez-Martinez:2016lgt,Blennow:2016jkn}. If the UV effects are parametrized by the $\alpha$ parameters~\cite{Escrihuela:2015wra}, they are constrained to be $\lsim 10^{-3}$, or smaller~\cite{Blennow:2016jkn}, and we should obtain the $\xi$ parameter bound of the similar order. Thus, our unitary method works in the presence of high-scale UV. 

The low-scale UV scenarios may be described by using the system of three active plus generic $N$ sterile neutrinos, the model known since the early days, see e.g. Refs.~\cite{Schechter:1980gr,Barger:1980tfa}. Within this framework, a sterile sector model-independent description of low-scale UV is attempted~\cite{Fong:2016yyh,Fong:2017gke}. In such scenarios the bounds on the $\alpha$ parameters are milder by at least one order of magnitude, and even more milder for $\alpha_{\tau \tau}$. See e.g., Refs.~\cite{Blennow:2016jkn,Parke:2015goa,Ellis:2020hus}. A rough estimation in Appendix~\ref{sec:constraint} reveals the current upper bound on $\xi$ of about 0.1 or less. Clearly, we need the better bound on $\xi$ to ensure the validity of our unitarity method.  

\section{Determination of oscillation probability without $\nu$SM ansatz} 
\label{sec:model-indep}

In the rest of this paper, we proceed with assumption of no unitarity violation in the three active neutrino space until Appendix~\ref{sec:constraint}.
To put the unitarity method for measuring $P(\nu_{\mu} \rightarrow \nu_{\tau})$ into practice we need to determine the neutrino oscillation probabilities $P(\nu_{\mu} \rightarrow \nu_{\mu})$ and $P(\nu_{\mu} \rightarrow \nu_{e})$. As we have learnt in section~\ref{sec:unitarity-method} our unitarity method does not necessitate the three-flavor $\nu$SM ansatz, we want to carry this task out in the theoretical-model independent way, as much as possible. 

To give this general idea a concrete shape, we present a toy analysis in this section assuming the experimental setting of the LBL accelerator neutrino experiment with muon neutrino beam. Since analyses of the atmospheric neutrino data are quite involved we focus on accelerator LBL experiment from now on. Among the two ongoing LBL experiments, T2K~\cite{T2K:2019bcf} and NOvA~\cite{NOvA:2019cyt}, we focus on the latter because of its higher energy beam. While exploration of $\tau$ neutrino physics using the CC $\tau$ production may require a higher energy neutrino beam, we should wait for LBNF nominal, or preferably its $\tau$-optimized configurations~\cite{Machado:2020yxl} for this purpose. We will be merited by the fact that NOvA has the functionally identical near and far detectors: A large fraction of the systematic errors would cancel between the two detectors. 

To show the basic idea of our toy analysis, we assume the quasi-elastic CC reactions $\nu_{e} N \rightarrow e^- N^{\prime}$ and $\nu_{\mu} N \rightarrow \mu^- N^{\prime}$ for detection of $\nu_{e}$ and $\nu_{\mu}$ at both the near and far detectors. The choice, where $N$ and $N^{\prime}$ denote, respectively, the target and produced nuclei, enables us to reconstruct the initial neutrino energy $E_{\nu}$ via the two-body kinematics. Nonetheless, by using the data in which the events with four hadronic energy-fraction quartiles~\cite{NOvA:2018gge,Vinton:2018aqq} are added, purity of the quasi-elastic nature of the CC events sample may be slightly harmed. Yet, we hope that the major part of this and the related problems are taken care of by the resultant relatively large error bars possessed by the results of $P(\nu_{\mu} \rightarrow \nu_{\mu})$ and $P(\nu_{\mu} \rightarrow \nu_{e})$ obtained by our method. 

To describe the principle of our analysis, we hereafter discuss explicitly only the neutrino channels, but the way how the antineutrino channels can be handled should be obvious from the neutrino channel discussion. After a brief description of event number distribution via the quasi-elastic CC reactions in section~\ref{sec:event-number}, we carry out our analyses for $P(\nu_{\mu} \rightarrow \nu_{\mu})$ and $P(\nu_{\mu} \rightarrow \nu_{e})$ in sections~\ref{sec:P-mumu} and \ref{sec:P-mue}, respectively. Then, we obtain $P(\nu_{\mu} \rightarrow \nu_{\tau})$ by our unitarity method in section~\ref{sec:P-mu-tau}. The similar analyses for the antineutrino channel probabilities will be repeated in section~\ref{sec:P-antinu}. 

\subsection{Theoretical expression of the event number distribution} 
\label{sec:event-number}

Muon neutrinos $\nu_{\mu}$ of energy $E_\nu$ in the neutrino beam reach a detector at distance $L$ from the production point as $\nu_{\mu}$ or $\nu_{e}$ with the probabilities $P(\nu_{\mu} \rightarrow \nu_{\mu}: E_\nu, L )$ and $P(\nu_{\mu} \rightarrow \nu_{e}: E_\nu, L )$, respectively. The event number distribution at the detector by the CC reaction $\nu_{\mu} N \rightarrow \ell_{\alpha}^- N^{\prime}$, where $\ell_{\alpha}$ ($\alpha = e, \mu, \tau$) are $SU(2)_L$ doublet, can be written as a function of neutrino energy $E_\nu$ as 
\begin{eqnarray} 
&& 
\frac{ d N_{\ell_\alpha} }{ d E_\nu } (\nu_{\mu} N \rightarrow \ell_{\alpha}^- N^{\prime}: L ) 
\nonumber \\
&=&
N_{T} \Phi_{\nu_\mu} ( E_\nu, L )
P(\nu_{\mu} \rightarrow \nu_{\alpha} : E_\nu, L ) 
\int d E_{\ell_\alpha} 
\epsilon ( E_{\ell_\alpha} ) \frac{ d \sigma }{ d E_{\ell_\alpha} } 
+ B_{\alpha} (E_\nu, L), 
\nonumber \\
&\equiv& 
S_{\alpha} ( E_\nu, L ) 
P(\nu_{\mu} \rightarrow \nu_{\alpha} : E_\nu, L ) 
+ B_{\alpha} (E_\nu, L), 
\label{dN-dEnu-alpha}
\end{eqnarray}
where $N_T$ denotes the number of target particles, $\Phi_{\nu_\mu} ( E_\nu, L )$ is the $\nu_\mu$ flux at the distance $L$ from the source, and $\frac{ d \sigma }{ d E_{\ell_\alpha} }$ is the cross section of the CC reaction $\nu_{\mu} N \rightarrow \ell_{\alpha}^- N^{\prime}$. which produces $\ell_\alpha$ lepton of energy $E_{\ell_\alpha}$. $\epsilon ( E_{\ell_\alpha} )$ denotes energy-dependent efficiency of identifying $\ell_\alpha$ lepton. In the last line of Eq.~\eqref{dN-dEnu-alpha} $S_{\alpha} ( E_\nu, L )$ and $B_{\alpha} (E_\nu, L)$ denote, respectively, the contributions from signal events without oscillation, and from background events.

As it stands, the expression in Eq.~\eqref{dN-dEnu-alpha} does not fully respect the experimental reality. The energy of neutrinos that undergo the CC reactions must be reconstructed using the reaction kinematics, and $E_{\nu}$ in Eq.~\eqref{dN-dEnu-alpha} must be understood as the reconstructed energy. In this process the various issues, e.g., the detector energy resolution and the effect of Fermi motion (as the target nucleus is in nuclei) have to be taken into account. Equation~\eqref{dN-dEnu-alpha} assumes that the error associated with this reconstruction process is small compared to the genuine neutrino energy. The assumption seems to be supported by the result of simulation which reports less than 10\% error in the reconstructed energy~\cite{Vinton:2018aqq}. See section~\ref{sec:limitations} for a brief description of how Eq.~\eqref{dN-dEnu-alpha} may be justified. 
A final comment on Eq.~\eqref{dN-dEnu-alpha} is that the sum over the various CC reactions, $\sum_{a} \epsilon_{a} ( E_{\mu} ) \frac{ d \sigma_{a} }{ d E_{\mu} }$ where $a$ denotes indices for the varying reaction channels, must be introduced with varying efficiencies. As it can be done without affecting the validity of our following discussion, we keep our simple expression Eq.~\eqref{dN-dEnu-alpha} as it is, with understanding that the summing over the CC reactions is always meant. 

Despite these and possibly other drawbacks, we use the expression in Eq.~\eqref{dN-dEnu-alpha} as the toy model for the event number distribution as a function of reconstructed neutrino energy. Despite that we do not write down the explicit expressions of the similar formulas in the antineutrino channels, they are easily obtained in an analogous way as in the neutrino channels. 

A few comments on the NOvA data used in this paper: From start to almost the end of our analysis we have consulted and used the information given in Ref.~\cite{NOvA:2019cyt}, which is then updated in Ref.~\cite{Himmel-Nu2020}. Very recently a new paper appeared from the NOvA collaboration~\cite{NOvA:2021nfi} which reports all the available data to date in the neutrino and antineutrino channels. It appears that the Monte Carlo analysis code is completely renewed. In each period, the data are conveniently made available at the NOvA data release~\cite{NOvA-data}, and we utilize the most recent version of it in our analysis.  

\subsection{Determination of disappearance probability $P(\nu_{\mu} \rightarrow \nu_{\mu})$}
\label{sec:P-mumu}

Now, we describe a method for extracting the survival (or disappearance) probability $P(\nu_{\mu} \rightarrow \nu_{\mu})$. Quite conveniently for our purpose, the experimental groups not only provide the experimental data of 
$\frac{ d N_{e} }{ d E_\nu } (\nu_{\mu} N \rightarrow \mu^- N^{\prime}: L_{ \text{far} })$, the left-hand-side (LHS) of Eq.~\eqref{dN-dEnu-alpha} ($\alpha = \mu$), but also Monte Carlo expectation of the same quantity without oscillation. If we take the ratio of these two quantities at the far detector distance, we obtain 
\begin{eqnarray} 
&& 
\frac{ \frac{ d N_{\mu} }{ d E_\nu } (\nu_{\mu} N \rightarrow \mu^- N^{\prime}: L_{ \text{far} }) } 
{ \frac{ d N_{\mu} }{ d E_\nu } (\nu_{\mu} N \rightarrow \mu^- N^{\prime}: L_{ \text{far} }) \vert_{ \text{no oscillation} } } 
=
\frac{ P(\nu_{\mu} \rightarrow \nu_{\mu} : E_\nu, L_{ \text{far} } ) + r_{\mu} ( L_{ \text{far} } ) }{ 1 + r_{\mu} ( L_{ \text{far} } ) },
\label{P-mumu}
\end{eqnarray}
where we have defined the background to signal ratio 
\begin{eqnarray} 
&& 
r_{\alpha} 
\equiv 
\frac{ B_{\alpha} (E_\nu, L) }{ S_{\alpha} ( E_\nu, L ) } ~~~~~~
(\alpha = e, \mu). 
\label{r-def}
\end{eqnarray}
The right-hand side (RHS) of Eq.~\eqref{P-mumu} is almost the probability, apart from the $r_{\mu}$ terms, because all the factors other than these cancel out between the numerator and the denominator. 
This cancellation takes place even in the case that sum over the varying reaction channels are introduced in the CC reactions to produce muons, as mentioned earlier. It may be relevant for higher hadronic energy-fraction quartiles~\cite{NOvA:2018gge,Vinton:2018aqq}. 

Thanks to the experimental group the information of the background is also provided~\cite{NOvA-data}, and hence we can obtain the disappearance probability $P(\nu_{\mu} \rightarrow \nu_{\mu})$. The background for $\nu_{\mu}$ disappearance CC events is about 4\% level for neutrino and 3\% level for antineutrino channels, respectively~\cite{NOvA:2019cyt}. 
Notice that in plotting the event number distribution as a function of reconstructed neutrino energy, the effects of energy smearing through the event reconstruction process as well as by the Fermi motion are taken care of by the experimental group. The same comment applies to the plot for extracting $P(\nu_{\mu} \rightarrow \nu_{e})$ in section~\ref{sec:P-mue}. 

\begin{figure}[h!]
\begin{center}
\vspace{3mm}
\includegraphics[width=0.70\textwidth]{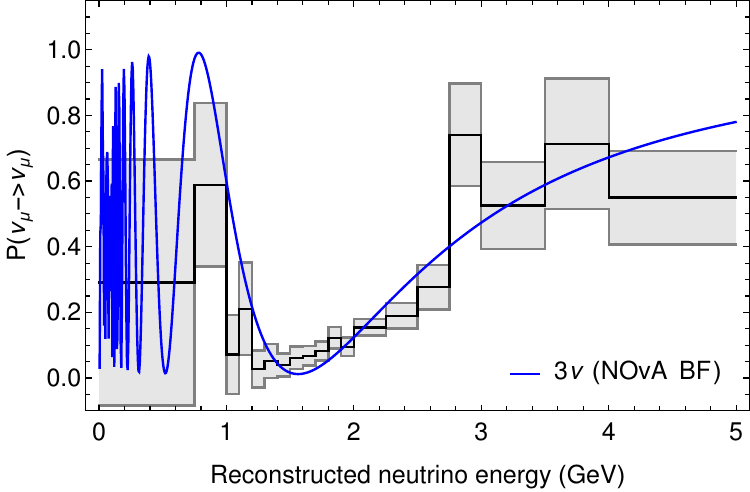}
\end{center}
\vspace{-3mm}
\caption{
The disappearance probability $P(\nu_{\mu} \rightarrow \nu_{\mu})$ calculated by using Eq.~\eqref{P-mumu} with the data given in the NOvA data release~\cite{NOvA-data} is shown by the thick black histogram and the 1$\sigma$ error band by the shaded gray region. 
The blue line shows the $\nu$SM oscillation probability calculated with the mixing parameters given in Table IV in Ref.~\cite{NOvA:2019cyt}. In the legend ``NOvA BF'' stands for the ``NOvA best fit''. 
} 
\vspace{-2mm}
\label{fig:P-mumu}
\end{figure}

Therefore, the determination of $P(\nu_{\mu} \rightarrow \nu_{\mu})$ through Eq.~\eqref{P-mumu} would be the cleanest way among the methods for determining the oscillation probability we discuss in this paper. Notice that our method is a data-driven way, and we do not rely on the expression of the oscillation probability calculated by the $\nu$SM standard three-flavor oscillation. The obtained result for $P(\nu_{\mu} \rightarrow \nu_{\mu})$ is presented in Fig.~\ref{fig:P-mumu} with the black histogram and its 1$\sigma$ error band as the shaded gray region. 

The blue line in Fig.~\ref{fig:P-mumu} is the $\nu$SM three-neutrino expression of $P(\nu_{\mu} \rightarrow \nu_{\mu})$ with the mixing parameters used by the NOvA group, in Table IV in Ref.~\cite{NOvA:2019cyt}. For simplicity and brevity we call this parameter set as the ``NOvA best fit''. Figure~\ref{fig:P-mumu} indicates that the obtained result of $P(\nu_{\mu} \rightarrow \nu_{\mu})$ is consistent with the standard three neutrino oscillation. In fact, the $\nu$SM line passes through the 1$\sigma$ uncertainty band of the obtained histogram in almost all the bins.  
As mentioned above, and will be further discussed in section~\ref{sec:independence}, our method for determining $P(\nu_{\mu} \rightarrow \nu_{\mu})$ does not rely on the $\nu$SM. Therefore, we do not judge whether our method is successful or not by how close our result is to the $\nu$SM. 

In certain limited energy regions in Fig.~\ref{fig:P-mumu}, the 1$\sigma$ error band of $P(\nu_{\mu} \rightarrow \nu_{\mu})$ penetrates into the unphysical regions of $P(\nu_{\mu} \rightarrow \nu_{\mu}) < 0$. Similarly, later in section~\ref{sec:P-antinu} 
we will see that $P(\bar{\nu}_{\mu} \rightarrow \bar{\nu}_{\mu})$ expands into the region $> 1$. See Fig.~\ref{fig:anti-Pmumu-Pmue}. We expect that these features will disappear as the better statistics of events is accumulated. 

\subsection{Determination of appearance probability $P(\nu_{\mu} \rightarrow \nu_{e})$}
\label{sec:P-mue}

Now, we discuss determination of $P(\nu_{\mu} \rightarrow \nu_{e}: E_\nu, L_{ \text{far} } )$. In our simple-minded experimental setting of the LBL neutrino experiment we assume that the near detector is placed at a location so close to the neutrino production point such that one can safely assume that $P(\nu_{\mu} \rightarrow \nu_{\mu}: L_{ \text{near} }) =1$. This is a good approximation for the NOvA experiment because $L_{ \text{near} } / L_{ \text{far} } \approx 10^{-3}$. Then, the relevant ratio of the event number distributions is given, by using Eq.~\eqref{dN-dEnu-alpha}, as\footnote{
The notation $\frac{ d N_{e} }{ d E_\nu } (\nu_{\mu} N \rightarrow e^- N^{\prime}: L_{ \text{far} })$ may be a little confusing because it hides the process of $\nu_{\mu}$ to $\nu_{e}$ oscillation and the subsequent CC reaction $\nu_{e} N \rightarrow e^- N^{\prime}$. But, it is the notation we have defined in Eq.~\eqref{dN-dEnu-alpha}.
}
\begin{eqnarray} 
&& 
\frac{ \frac{ d N_{e} }{ d E_\nu } (\nu_{\mu} N \rightarrow e^- N^{\prime}: L_{ \text{far} }) }
{ \frac{ d N_{\mu} }{ d E_\nu } (\nu_{\mu} N \rightarrow \mu^- N^{\prime}: L_{ \text{near} }) } 
= 
f_{e \mu}
\frac{  
P(\nu_{\mu} \rightarrow \nu_{e}: E_\nu, L_{ \text{far} } ) 
+ r_{e} ( L_{ \text{far} }) }
{ 1 + r_{\mu} ( L_{ \text{near} }) }, 
\label{ratio-e-mu} 
\end{eqnarray}
where $r_{e}$ and $r_{\mu}$ are defined in Eq.~\eqref{r-def}, and $f_{e \mu}$ is defined by 
\begin{eqnarray} 
&&
f_{e \mu} 
\equiv 
\frac{ S_{e} ( E_\nu, L_{ \text{far} } ) }{ S_{\mu} ( E_\nu, L_{ \text{near} } ) } 
= 
\frac{ N_{T}^{ \text{far} } \Phi_{\nu_\mu} ( E_\nu, L_{ \text{far} } ) 
\int _{E_0 }^{ E_\nu } d E_{e} \epsilon ( E_{e} )_{ \text{far} }  \frac{ d \sigma }{ d E_{e} } }
{ N_{T}^{ \text{near} } \Phi_{\nu_\mu} ( E_\nu, L_{ \text{near} } ) 
\int _{E_0 }^{ E_\nu } d E_{\mu} \epsilon ( E_{\mu} )_{ \text{near} } \frac{ d \sigma }{ d E_{\mu} }  }. 
\label{f-emu-def}
\end{eqnarray}
The ratio $f_{e \mu}$ is the far-to-near flux ratio weighted by (i) the detector volumes and (ii) the efficiencies averaged over the event number distributions. 
Since the LHS of Eq.~\eqref{ratio-e-mu}, the both numerator and denominator, is given by the experimental group, we can determine the appearance probability $P(\nu_{\mu} \rightarrow \nu_{e}: E_\nu, L_{ \text{far} } )$ if we know $f_{e \mu}$, $r_{e} ( L_{ \text{far} })$, and $r_{\mu} ( L_{ \text{near} })$.

\begin{figure}[h!]
\begin{center}
\vspace{1mm}
\includegraphics[width=0.7\textwidth]{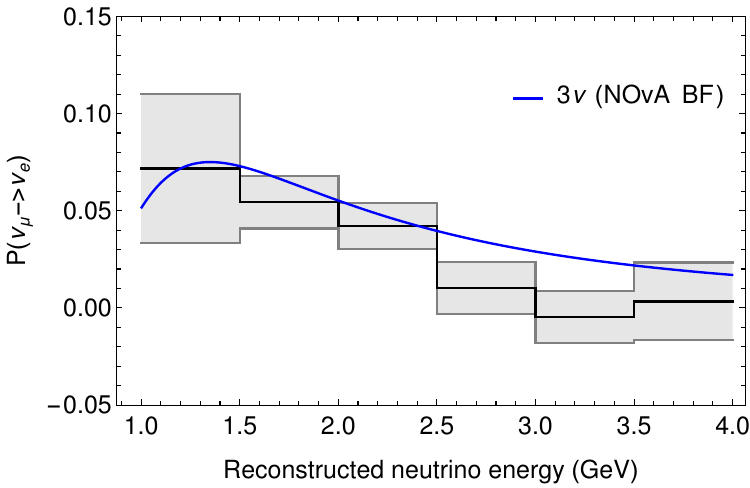}
\end{center}
\caption{  
Plotted by the thick black histogram is the appearance probability $P(\nu_{\mu} \rightarrow \nu_{e}: E_\nu, L_{ \text{far} } )$ calculated by using \eqref{P-mue-data}, and the shaded gray region is its 1$\sigma$ error. 
The blue line shows the $\nu$SM oscillation probability calculated with the mixing parameters given in Table IV in Ref.~\cite{NOvA:2019cyt}. In the legend ``NOvA BF'' stands for the ``NOvA best fit''. 
} 
\vspace{-2mm}
\label{fig:P-mue}
\end{figure}

Despite that the experimental group keeps the information on $f_{e \mu}$ not public, the result of Monte Carlo calculation is given for the event number distribution of electrons at the far detector in Slide 23 of Ref.~\cite{Himmel-Nu2020} (see Fig.4 in Ref.~\cite{NOvA:2021nfi}): 
\begin{eqnarray} 
&& 
\frac{ d N_{e} }{ d E_\nu } (\nu_{\mu} N \rightarrow e^- N^{\prime}: L_{ \text{far} }) \vert_{MC}
= 
f_{e \mu} 
P(\nu_{\mu} \rightarrow \nu_{e}: E_\nu, L_{ \text{far} } ) \vert_{MC} ~
\frac{ d N_{\mu} }{ d E_\nu } (\nu_{\mu} N \rightarrow \mu^- N^{\prime}: L_{ \text{near} }) 
%
\label{dN-dE-MC}
\end{eqnarray}
where $\frac{ d N_{\mu} }{ d E_\nu } (\nu_{\mu} N \rightarrow \mu^- N^{\prime}: L_{ \text{near} })$ is given by the NOvA experimental measurement and $P(\nu_{\mu} \rightarrow \nu_{e}: E_\nu, L_{ \text{far} } ) \vert_{MC}$ is calculated by using the ``NOvA best fit''. Quite conveniently, the both quantities are included in the NOvA data release~\cite{NOvA-data}. Then, one can solve Eq.~\eqref{dN-dE-MC} for $f_{e \mu}$. 

Now, we must note that $f_{e \mu}$ defined in \eqref{f-emu-def} contains only the information on the signal events, not background. Therefore, to evaluate $f_{e \mu}$ by using Eq.~\eqref{dN-dE-MC} we must restrict both LHS and RHS of Eq.~\eqref{dN-dE-MC} the information of the signal events only. One can easily satisfy this condition for the quantities obtained by MC, but not the muon number distribution, the last factor in Eq.~\eqref{dN-dE-MC} because it is the data. But, this problem is easily avoided if one insert everything into Eq.~\eqref{ratio-e-mu}, which entails 
\begin{eqnarray} 
&& 
\frac{ d N_{e} }{ d E_\nu } (\nu_{\mu} N \rightarrow e^- N^{\prime}: L_{ \text{far} })
=
\frac{ \frac{ d N_{e} }{ d E_\nu } (\nu_{\mu} N \rightarrow e^- N^{\prime}: L_{ \text{far} }) \vert_{MC} }
{ P(\nu_{\mu} \rightarrow \nu_{e}: E_\nu, L_{ \text{far} } ) \vert_{MC}  } 
\frac{  
P(\nu_{\mu} \rightarrow \nu_{e}: E_\nu, L_{ \text{far} } ) 
+ r_{e} ( L_{ \text{far} }) }
{ 1 + r_{\mu} ( L_{ \text{near} }) }.
%
\label{P-mue-data} 
\end{eqnarray}
The obtained result of $P(\nu_{\mu} \rightarrow \nu_{e}: E_\nu, L_{ \text{far} } )$ is presented in Fig.~\ref{fig:P-mue}. Again the $\nu$SM blue line is consistent with $P(\nu_{\mu} \rightarrow \nu_{e})$ in Fig.~\ref{fig:P-mue}. 

As in the case of $P(\nu_{\mu} \rightarrow \nu_{\mu})$ in Fig.~\ref{fig:P-mumu}, $P(\nu_{\mu} \rightarrow \nu_{e}: E_\nu, L_{ \text{far} } )$ goes into unphysical regions. Even the think black line ($3.0 \leq E_\nu \leq 3.5$ GeV) as well as the lower end of gray-shaded region ($2.5 \leq E_\nu \leq 4.0$ GeV) go down into minus. We expect that these features will disappear as the measurements further proceed. 

\subsection{$\nu$ Standard Model independence of our method and its significance in wider contexts}
\label{sec:independence}

Now some of the readers may argue that by using the NOvA Monte Carlo simulation results in Eq.~\eqref{P-mue-data} our analysis depend on the standard three-flavor model of oscillation. If so, we can no longer claim that it is independent of the $\nu$SM paradigm. Fortunately, this is not the case. Notice that $\frac{ d N_{e} }{ d E_\nu } (\nu_{e} N \rightarrow e^- N^{\prime}: L_{ \text{far} }) \vert_{MC}$ scales as $P(\nu_{\mu} \rightarrow \nu_{e}: E_\nu, L_{ \text{far} }) \vert_{MC}$ apart from the small background contributions. Then, the dependence on $P(\nu_{\mu} \rightarrow \nu_{e}: E_\nu, L_{ \text{far} }) \vert_{MC}$ cancels between the numerator and the denominator in Eq.~\eqref{P-mue-data}, allowing us to remain essentially in the $\nu$SM independent analysis. 

Extraction of the oscillation probabilities $P(\nu_{\mu} \rightarrow \nu_{e})$ and $P(\nu_{\mu} \rightarrow \nu_{\mu})$ in a model-independent way may be important in much wider contexts beyond the unitarity method for $P(\nu_{\mu} \rightarrow \nu_{\tau})$. 
Currently the experimental results are reported by showing the best-fit values of the mixing angles and the CP phase by assuming the $\nu$SM parametrization. While it is a valid way, the result is of course $\nu$SM dependent. Instead, a model-independent extraction of the oscillation probability itself could directly signal effects outside the $\nu$SM. It can be done immediately with the currently available data, but it would become an indispensable alternative in high-statistics experiments like T2HK and DUNE. 

\section{Determination of $P(\nu_{\mu} \rightarrow \nu_{\tau})$} 
\label{sec:P-mu-tau} 

\begin{figure}[h!]
\begin{center}
\vspace{3mm}
\includegraphics[width=0.7\textwidth]{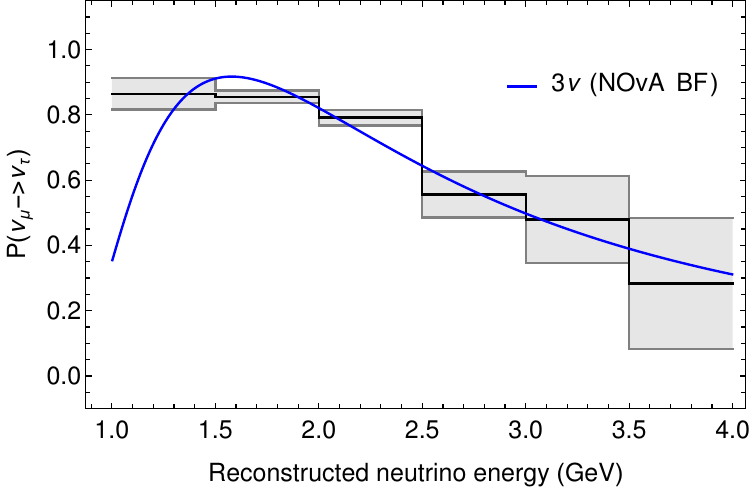}
\includegraphics[width=0.7\textwidth]{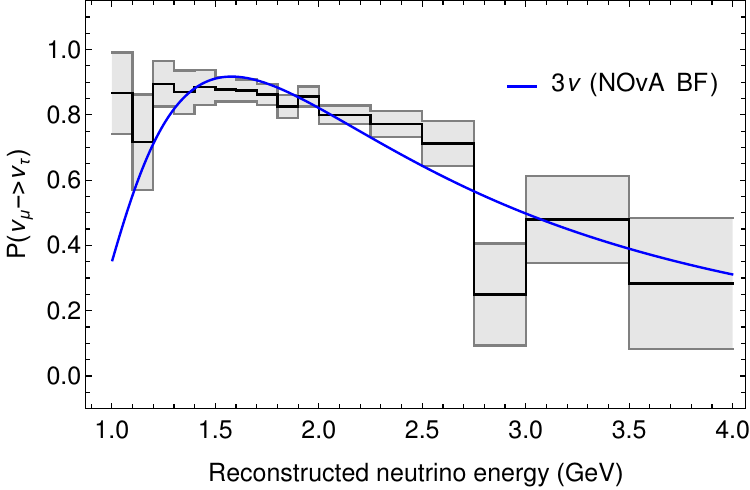}
\end{center}
\caption{  
The $\tau$ neutrino appearance probability $P(\nu_{\mu} \rightarrow \nu_{\tau}: E_\nu, L_{ \text{far} } )$ is plotted by the thick black histogram, which is calculated by using unitarity with the $\nu_{\mu} \rightarrow \nu_{\mu}$ and $\nu_{\mu} \rightarrow \nu_{e}$ probabilities in Figs.~\ref{fig:P-mumu} and \ref{fig:P-mue}, respectively. The gray shaded area is its 1$\sigma$ error. 
In the upper panel, to combine $P(\nu_{\mu} \rightarrow \nu_{\mu})$ and $P(\nu_{\mu} \rightarrow \nu_{e})$, we take the coarse bins for the both channels. In the lower panel, we have kept the original bin sizes of the both $P(\nu_{\mu} \rightarrow \nu_{\mu})$ and $P(\nu_{\mu} \rightarrow \nu_{e})$ as in Figs.~\ref{fig:P-mumu} and ~\ref{fig:P-mue}, respectively. See the text for more details. The blue line shows the $\nu$SM oscillation probability calculated with the ``NOvA best fit''.
} 
\vspace{-2mm}
\label{fig:P-mutau}
\end{figure}

Given our estimates of $P(\nu_{\mu} \rightarrow \nu_{\mu})$ and $P(\nu_{\mu} \rightarrow \nu_{e})$ in Figs.~\ref{fig:P-mumu} and \ref{fig:P-mue} in sections~\ref{sec:P-mumu} and \ref{sec:P-mue}, respectively, it is now straightforward to obtain $P(\nu_{\mu} \rightarrow \nu_{\tau})$ by using unitarity~\eqref{unitarity-relation}. The result is given in Fig.~\ref{fig:P-mutau}. The errors of $P(\nu_{\mu} \rightarrow \nu_{\mu})$ and $P(\nu_{\mu} \rightarrow \nu_{e})$ are added in quadrature. 
As before, the blue line shows the $\nu$SM oscillation probability calculated with the mixing parameters given in Table IV in Ref.~\cite{NOvA:2019cyt}, the ``NOvA best fit''.

We need to make some comments on Fig.~\ref{fig:P-mutau}, because we have presented the two panels. They differ in the binning, mainly at low energies $E \lsim 3$ GeV. The issue is that while $P(\nu_{\mu} \rightarrow \nu_{\mu})$ is determined with finer bins as seen in Fig.~\ref{fig:P-mumu}, $P(\nu_{\mu} \rightarrow \nu_{e})$ has coarse bins as in Fig.~\ref{fig:P-mue}. If we use the coarse bins for the both $P(\nu_{\mu} \rightarrow \nu_{\mu})$ and $P(\nu_{\mu} \rightarrow \nu_{e})$, the result in the upper panel is obtained. 
But, since $P(\nu_{\mu} \rightarrow \nu_{\mu})$ is much larger than $P(\nu_{\mu} \rightarrow \nu_{e})$ in most bins, we could combine $P(\nu_{\mu} \rightarrow \nu_{\mu})$ and $P(\nu_{\mu} \rightarrow \nu_{e})$ in such a way that the respective binning of $P(\nu_{\mu} \rightarrow \nu_{\mu})$ and $P(\nu_{\mu} \rightarrow \nu_{e})$ are kept as they are. If we take this attitude the obtained result of $P(\nu_{\mu} \rightarrow \nu_{\tau})$ is presented in the lower panel. 

A problem in our treatment for the lower panel $P(\nu_{\mu} \rightarrow \nu_{\tau})$ is, of course, we have to assume that $P(\nu_{\mu} \rightarrow \nu_{e})$ is constant over the energy regions of e.g., $E=1.0 - 1.5$ GeV, or $E=1.5 - 2$ GeV, whereas $P(\nu_{\mu} \rightarrow \nu_{\mu})$ changes in the region. Nonetheless, $P(\nu_{\mu} \rightarrow \nu_{\mu})$ significantly varies in the region $E=1.0 - 2.0$ GeV, so that keeping the information with finer bins would make sense. These are the reasonings for which we wind up to present the two panels in Fig.~\ref{fig:P-mutau}. The blue line for the $\nu$SM three-neutrino expression of $P(\nu_{\mu} \rightarrow \nu_{\tau})$ reasonably fit to our results both in the upper and lower panels. 

The $\lsim5$\% measurement of $P(\nu_{\mu} \rightarrow \nu_{\tau})$ around the peak region $1.5 < E < 2.5$ GeV reported in Fig.~\ref{fig:P-mutau} is certainly intriguing. But, we postpone our comment to section~\ref{sec:anti-P-mutau} where we make comparison between the results of $P(\nu_{\mu} \rightarrow \nu_{\tau})$ in the neutrino and antineutrino ($\nu \rightarrow \bar{\nu}$) channels.  

\section{Determination of probabilities in the antineutrino channels}
\label{sec:P-antinu} 

In this section, we repeat the same exercise for the antineutrino channels, the ones we have carried out in sections~\ref{sec:model-indep} and \ref{sec:P-mu-tau} for the neutrino channels. The antineutrino channels are important to obtain the information on CP violation in combination with the neutrino channel. 

\begin{figure}[h!]
\begin{center}
\includegraphics[width=0.7\textwidth]{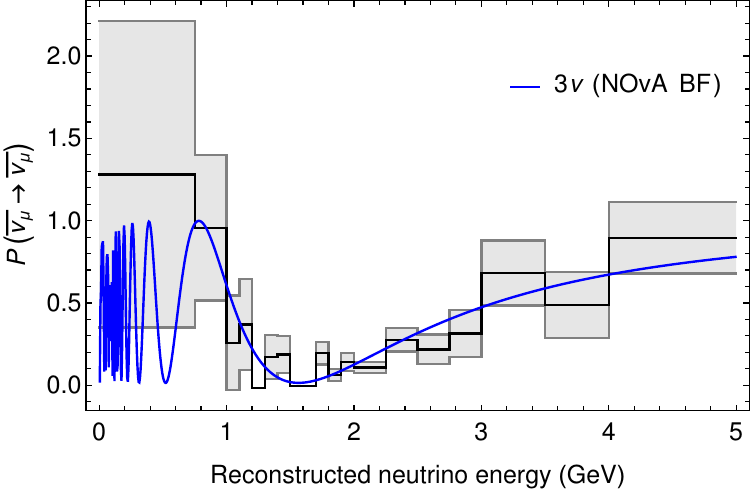}
\includegraphics[width=0.7\textwidth]{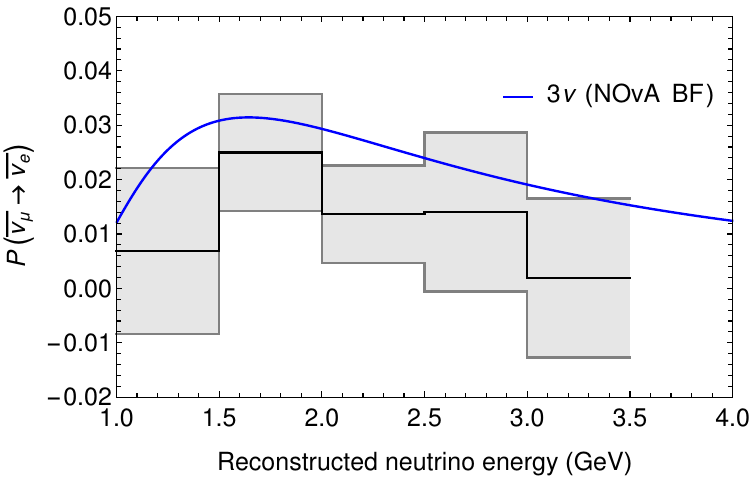}
\end{center}
\caption{  
Plotted by the think black histograms are the disappearance probability 
$P(\bar{\nu}_{\mu} \rightarrow \bar{\nu}_{\mu}: E_\nu, L_{ \text{far} } )$ (upper panel), 
and the appearance probability 
$P(\bar{\nu}_{\mu} \rightarrow \bar{\nu}_{e}: E_\nu, L_{ \text{far} } )$ (lower panel), which are calculated by using the antineutrino versions of Eqs.~\eqref{P-mumu} and \eqref{P-mue-data}, respectively, and their 1$\sigma$ errors are shown by the shaded gray regions. 
The blue line shows the $\nu$SM oscillation probability calculated with the ``NOvA best fit''. 
} 
\vspace{-2mm}
\label{fig:anti-Pmumu-Pmue}
\end{figure}

\begin{figure}[h!]
\begin{center}
\vspace{1mm}
\includegraphics[width=0.7\textwidth]{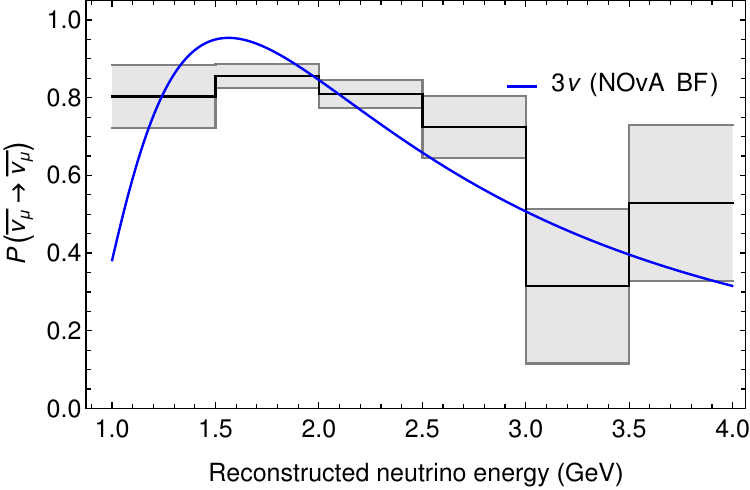}
\includegraphics[width=0.7\textwidth]{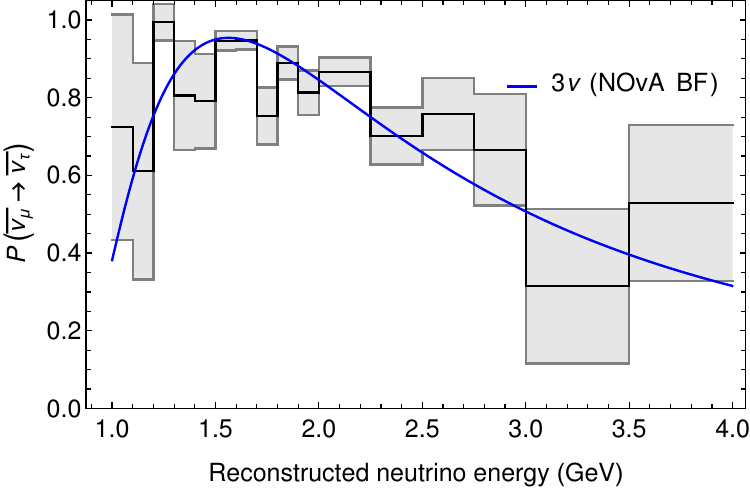}
\end{center}
\caption{ 
The antineutrino appearance probability $P(\bar{\nu}_{\mu} \rightarrow \bar{\nu}_{\tau}: E_\nu, L_{ \text{far} } )$ and its 1$\sigma$ error region are presented with the same style as in Fig.~\ref{fig:P-mutau} for the neutrino version. 
The blue line shows the $\nu$SM oscillation probability calculated with the ``NOvA best fit''.
} 
\vspace{-2mm}
\label{fig:anti-Pmutau}
\end{figure}

In Fig.~\ref{fig:anti-Pmumu-Pmue}, plotted is the disappearance probability $P(\bar{\nu}_{\mu} \rightarrow \bar{\nu}_{\mu}: E_\nu, L_{ \text{far} } )$ (upper panel), and the appearance probability $P(\bar{\nu}_{\mu} \rightarrow \bar{\nu}_{e}: E_\nu, L_{ \text{far} } )$ (lower panel), which are calculated by using the antineutrino versions of Eqs.~\eqref{P-mumu} and \eqref{P-mue-data}, respectively, and their 1$\sigma$ errors. The $\nu$SM oscillation probability calculated with the ``NOvA best-fit'' is also shown. 
Roughly speaking, the uncertainties in determination of 
$P(\bar{\nu}_{\mu} \rightarrow \bar{\nu}_{\mu}: E_\nu, L_{ \text{far} } )$ and $P(\bar{\nu}_{\mu} \rightarrow \bar{\nu}_{e}: E_\nu, L_{ \text{far} } )$ are comparable to each other. However, in the disappearance channels bin-to-bin fluctuations look somewhat larger in the antineutrino channel with a few vanishing number of event bins at around the oscillation maximum, i.e., the oscillation minimum in the disappearance channels $P(\bar{\nu}_{\mu} \rightarrow \bar{\nu}_{\mu})$. 

Probably due to lack of statistics the probability exceeds unity in a few low and high energy bins of $P(\bar{\nu}_{\mu} \rightarrow \bar{\nu}_{\mu}: E_\nu, L_{ \text{far} } )$. Similarly, $P(\bar{\nu}_{\mu} \rightarrow \bar{\nu}_{e}: E_\nu, L_{ \text{far} } )$ goes into minus at the similar low and high energy bins. 

\subsection{Result of $P(\bar{\nu}_{\mu} \rightarrow \bar{\nu}_{\tau})$ and its comparison with $P(\nu_{\mu} \rightarrow \nu_{\tau})$} 
\label{sec:anti-P-mutau}

In Fig.~\ref{fig:anti-Pmutau}, the $\bar{\nu}_{\tau}$ appearance probability $P(\bar{\nu}_{\mu} \rightarrow \bar{\nu}_{\tau}: E_\nu, L_{ \text{far} } )$ and its 1$\sigma$ error are plotted. The style of presentation and line symbols are the same as before. They are calculated by using the antineutrino version of unitarity~\eqref{unitarity-relation}. The upper panel is for the case of common coarse bin as used in $P(\bar{\nu}_{\mu} \rightarrow \bar{\nu}_{e}: E_\nu, L_{ \text{far} } )$, while the lower panel is for use of different binning, the finer bin for $P(\bar{\nu}_{\mu} \rightarrow \bar{\nu}_{\mu}: E_\nu, L_{ \text{far} } )$, and the coarse bin for $P(\bar{\nu}_{\mu} \rightarrow \bar{\nu}_{e}: E_\nu, L_{ \text{far} } )$, as done in the lower panel of Fig.~\ref{fig:P-mutau} in the neutrino channel. 

By comparing between the obtained $\nu_{\tau}$ and $\bar{\nu}_{\tau}$ appearance probabilities in Fig.~\ref{fig:P-mutau} and Fig.~\ref{fig:anti-Pmutau}, one can say that (1) The uncertainties of the appearance probabilities are comparable but slightly larger in the antineutrino channel. (2) More visibly, the bin by bin fluctuations are larger in the antineutrino channel. 
The accuracy of $P(\nu_{\mu} \rightarrow \nu_{\tau})$ itself is quite good with less than 5\% (8\%) error in the peak region $1.5 < E < 2.5$ GeV in the neutrino (antineutrino) channel. One may say that the accuracy of 5\%, or 8\%, is a superb performance, but it is basically achieved by the experimental measurement of $P(\nu_{\mu} \rightarrow \nu_{\mu})$ and $P(\nu_{\mu} \rightarrow \nu_{e})$, and what is done by our analysis is to translate the accuracies to $P(\nu_{\mu} \rightarrow \nu_{\tau})$.
Whereas the central value as well as the error of the antineutrino probability considerably fluctuate bin by bin, but at less than $\pm 10$\% level for the central value in the same peak region as above. In the above we are referring the finer bin versions of $P(\nu_{\mu} \rightarrow \nu_{\mu})$ and $P(\bar{\nu}_{\mu} \rightarrow \bar{\nu}_{\mu})$. 

Here, we note a possible mechanism of error reduction for $P(\nu_{\mu} \rightarrow \nu_{\tau})$, which is characteristic to our unitarity method. First of all the effect of $P(\nu_{\mu} \rightarrow \nu_{e})$ is relatively minor, and hence we disregard it in this discussion. In the disappearance channels the statistics is high, and we could assume that the errors are well characterized as a relative, percent error. The peak region of the appearance channel $\nu_{\mu} \rightarrow \nu_{\tau}$ corresponds to the region where $P(\nu_{\mu} \rightarrow \nu_{\mu})$ is small, so that the error of $P(\nu_{\mu} \rightarrow \nu_{\mu})$ is also small. The small error, through unitarity, leads to the small error of $P(\nu_{\mu} \rightarrow \nu_{\tau})$ in its peak region. Since $P(\nu_{\mu} \rightarrow \nu_{\tau})$ is large in the peak region, its percent error is even smaller. If this is the qualitatively correct explanation, it is a new merit of the unitarity method. When much higher statistics are gained, a smaller percent error of $P(\nu_{\mu} \rightarrow \nu_{\tau})$ than $P(\nu_{\mu} \rightarrow \nu_{\mu})$'s would manifest in regions where the both probabilities are large. 

The smallness of the error might also be because the experimental errors are not taken into account to a sufficient level or the error correlations is important. On the other hand, large fluctuations in $P(\bar{\nu}_{\mu} \rightarrow \bar{\nu}_{\tau})$ seem to tell us that accumulating a better event statistics is necessary, which is not easy to achieve in the LBL neutrino experiments.  
Even though we have included the T2K data to our analysis, it would not improve so much the accuracy of our unitarity-reconstructed $P(\nu_{\mu} \rightarrow \nu_{\tau})$ because the T2K events mostly span lower energy region than NOvA's, $E_{\nu} \lsim 1$ GeV. In this sense these two LBL experiments are complementary with each other by covering the different energy regions.\footnote{
It should be remembered that if interests point to the $\nu_{\tau}$ appearance probability at low energies, one must attempt the similar analysis by using the T2K data. } 

\section{A few final remarks} 
\label{sec:final-remark}

We have described our unitarity method for determining $\nu_{\tau}$ appearance probability $P(\nu_{\mu} \rightarrow \nu_{\tau})$, and examined performance of the method by taking the concrete case of the NOvA experiment. We believe that our analysis method is reasonably set up to allow model-independent determination of the probabilities $P(\nu_{\mu} \rightarrow \nu_{\alpha})$ ($\alpha = e, \mu, \tau$), and the results are indeed sensible. But, there are limitations inherent to our method. 

\subsection{Assumptions and limitations of our analysis} 
\label{sec:limitations}

The most important approximation we have made in deriving our basic equation~\eqref{dN-dEnu-alpha} is that the error in reconstructing the neutrino energy is much smaller than the genuine neutrino energy. Without this assumption we cannot factorize the oscillation probabilities as in Eq.~\eqref{dN-dEnu-alpha}. The point may be illuminated by a toy-model expression of the event number distribution as a function of the reconstructed neutrino energy $E_{ \text{rec} }$ in the reaction $\nu_{\mu}$ is oscillated to $\nu_{\alpha}$, and $\nu_{\alpha}$ undergoes CC reaction $\nu_{\alpha} + N \rightarrow \ell_{\alpha} + N^{\prime}$, 
\begin{eqnarray} 
&& 
\frac{ d N_{\ell \alpha} }{ d E_{ \text{rec} } }
= 
N_{T} \int d E_\nu  
\Phi_{\nu_\mu} ( E_\nu ) 
P(\nu_{\mu} \rightarrow \nu_{\alpha}: E_\nu ) 
\frac{1}{ \sqrt{ 2 \pi \sigma^2 } } e^{ - \frac{ \left( E_{ \text{rec} } - E_\nu \right)^2 }{ 2 \sigma^2 }} 
\int d E_{\ell \alpha}
\epsilon ( E_{\ell \alpha} ) 
\frac{ d \sigma }{ d E_{\ell \alpha} } ( E_\nu, E_{\ell \alpha} ),
\nonumber \\
\label{E-rec-distribution}
\end{eqnarray}
where we have assumed the Gaussian shape of $E_\nu$ reconstruction error function. 
Under the limit $\sigma \ll E_\nu$ Eq.~\eqref{E-rec-distribution} reproduces Eq.~\eqref{dN-dEnu-alpha}. Fortunately, the detailed study in Ref.~\cite{Vinton:2018aqq} assures the smallness of the error in neutrino energy reconstruction to be less than 10\%, which is indeed small but not vanishingly small. 

Further ``limitation discussions'' on our analysis would entail an endless list. For example, mistreatment of error correlations, or double counting of the errors, etc. Or, one could raise the possibility of analysis without binning. 
We are reluctant to enter into the detailed discussions of these or the other points here.  It is because, we believe, improving our toy analysis is {\em not} the right way to proceed. What is really needed is the real analysis by the experimental group. 

\subsection{Improving the bound on non-unitarity}
\label{sec:UV-bound} 

Improving the constraints on non-unitarity, in our case on the $\xi$ parameter, is important to strengthen the basis of our unitarity method for $P(\nu_{\mu} \rightarrow \nu_{\tau})$. In more generic context including the $\xi$ bound, we expect that the better constraints on UV which improve the current ones~\cite{Blennow:2016jkn,Parke:2015goa,Ellis:2020hus} will be obtained before DUNE starts to do $\tau$ neutrino physics. It will be done, for example, by the ongoing and upcoming experiments such as SBN program at Fermilab~\cite{Machado:2019oxb}, JSNS2~\cite{JSNS2:2021hyk}, T2K~\cite{T2K:2019bcf}, NOvA~\cite{NOvA:2019cyt}, Super-K~\cite{Super-Kamiokande:2017yvm}, IceCube~\cite{Stuttard:2020zsj,IceCubeCollaboration:2021euf}, KM3NeT~\cite{KM3NeT:2021ozk}, JUNO~\cite{JUNO:2015zny,Fong:2016yyh}, and possibly Hyper-K~\cite{Hyper-Kamiokande:2018ofw}. 
These are the case of low-scale UV (or low mass sterile leptons) and the bound is already much severer in high-scale UV case, $\lsim 10^{-3}$~\cite{Blennow:2016jkn}. 

\subsection{Absolute neutrino flux} 
\label{sec:nu-flux}

If our purpose is restricted to determine the oscillation probability only, the necessity of knowing the precise muon neutrino flux may be relaxed because the near-far detector comparison basically does the job. The fact that NOvA has the functionally identical near and far detectors certainly helps. However, to measure $\nu_{\tau}$ CC cross sections with comparable accuracy with $\nu_{\mu}$'s, and to study possible new physics effects in the $\nu_{\tau}$ induced reactions, we would need to know the absolute neutrino flux, hopefully to the accuracy better than what are achieved for the ongoing projects~\cite{T2K:2012bge,MINERvA:2016iqn}. A method of using $\nu_{\mu} - e$ scattering is suggested~\cite{Marshall:2019vdy} based on the measurement in MINERvA~\cite{MINERvA:2019hhc}.

\section{Concluding remarks}
\label{sec:conclusion}

In this paper we have described a way of determining $\tau$ neutrino appearance probability $P(\nu_{\mu} \rightarrow \nu_{\tau})$ using unitarity in the $\nu$SM independent way. Despite our analysis is at the level of theorists' toy exercise, we hope, we were able to demonstrate the ``in principle feasibility'' of the unitarity method for measuring $P(\nu_{\mu} \rightarrow \nu_{\tau})$. Of course, the experimentalist-level real analysis must be performed to give the idea a realistic shape. If this paper acts as a trigger for this, it would be the most successful outcome of this paper. 

Once DUNE and Hyper-K turn on in the near future, we will enjoy the rich prospects for $\tau$ neutrino physics. They will carry out simultaneous measurement of $P(\nu_{\mu} \rightarrow \nu_{\mu})$ and $P(\nu_{\mu} \rightarrow \nu_{e})$, and $\nu_{\tau}$ appearance probability $P(\nu_{\mu} \rightarrow \nu_{\tau})$ can be determined by the unitarity method. Then, the promising expectation is that the oscillation-produced intense $\nu_{\tau}$ neutrino beam in DUNE can be used to investigate the properties of $\tau$ neutrino CC reactions. It is worth to note that all these processes take place in the DUNE experiment in an {\it in situ} manner. This feature would allow the reduction of systematic uncertainties by comparing between near and far detectors, and in mutual simultaneous analyses of the three observables. 
From the viewpoint of unitarity measurement of $P(\nu_{\mu} \rightarrow \nu_{\tau})$ from low to high energies (say, 400 MeV to $\sim$10 GeV), T2HK and DUNE will play complementary role as analogous to T2K-NOvA complementarity, but at much higher level of the accuracies. 

The method for measuring $P(\nu_{\mu} \rightarrow \nu_{\tau})$ via unitarity may be applicable to the atmospheric neutrino observation, because extraction of $P(\nu_{\mu} \rightarrow \nu_{\mu})$ and $P(\nu_{\mu} \rightarrow \nu_{e})$ from the data should be possible in the analyses.\footnote{
We thank Kimihiro Okumura for informative correspondences on the possible analysis of the atmospheric neutrino data of Super-K in the context of this paper. 
}
If it works in Super-K it will allow DUNE to enjoy the knowledge of $P(\nu_{\mu} \rightarrow \nu_{\tau})$ in the energy region of $1-10$ GeV from the first day of its operation. Notice that Super-K will be able to accumulate the atmospheric neutrino data for 30 years at the DUNE turn-on, which provide a rare, valuable chance of the international collaboration for tau neutrino physics. 

\appendix

\section{Constraints on unitarity violation parameter $\xi$} 
\label{sec:constraint}

We have introduced in section~\ref{sec:unitarity-method}, the $\xi$ parameter, a measure for UV, see eq.~\eqref{xi-def}. 
Let us make an order of magnitude estimation of $\xi$ by using the known constraints on non-unitarity. For this purpose we utilize the analytic formulas for the oscillation probabilities $\sum_{\beta = e, \mu, \tau} P(\nu_{\mu} \rightarrow \nu_{\beta})$ calculated to first order in the UV $\alpha$ parameters~\cite{Escrihuela:2015wra}. In Ref.~\cite{Martinez-Soler:2018lcy} we have formulated the perturbative framework serving for such purposes, started from the renormalized helio-perturbation theory~\cite{Minakata:2015gra} and perturb it by the $\alpha$ parameters. The resulting expressions of the oscillation probability $P(\nu_{\mu} \rightarrow \nu_{\beta})$ ($\beta = e, \mu, \tau$) has an interesting structure: 
\begin{eqnarray} 
&& 
P(\nu_{\mu} \rightarrow \nu_{\beta}) 
= 
P(\nu_{\mu} \rightarrow \nu_{\beta}) \vert_{\nu\text{SM} } 
+ P(\nu_{\mu} \rightarrow \nu_{\beta})_{ \text{ EV } }^{(1)} 
+ P(\nu_{\mu} \rightarrow \nu_{\beta})_{ \text{ UV } }^{(1)},
\label{P-mubeta} 
\end{eqnarray}
where the first term is the $\nu$SM part~\cite{Minakata:2015gra}, the second and third terms express the first-order corrections by the $\alpha$ parameters. Since $P(\nu_{\mu} \rightarrow \nu_{\beta}) \vert_{\nu\text{SM} }$ and the evolution part $P(\nu_{\mu} \rightarrow \nu_{\beta})_{ \text{ EV } }^{(1)}$ are unitary, 
$\sum_{\beta} P(\nu_{\mu} \rightarrow \nu_{\beta}) \vert_{\nu\text{SM} } = 1$ and 
$\sum_{\beta} P(\nu_{\mu} \rightarrow \nu_{\beta})_{ \text{ EV } }^{(1)} =0$, the $\xi$ parameter depends only on the unitarity violating correction $P(\nu_{\mu} \rightarrow \nu_{\beta})_{ \text{ UV } }^{(1)}$~\cite{Martinez-Soler:2018lcy}. Then, using the formulas given in Ref.~\cite{Martinez-Soler:2018lcy} we obtain $\xi$ to first order in the $\alpha$ parameters as
\begin{eqnarray} 
&& 
\xi (E) = 
- \left[ P(\nu_{\mu} \rightarrow \nu_{e})_{ \text{ UV } }^{(1)} 
+ P(\nu_{\mu} \rightarrow \nu_{\mu})_{ \text{ UV } }^{(1)} 
+ P(\nu_{\mu} \rightarrow \nu_{\tau})_{ \text{ UV } }^{(1)} \right] 
\nonumber \\
&& 
\hspace{-14mm} 
= 
- \sin 2\theta_{23} 
\biggl[
\mbox{Re} \left( e^{ i \delta } \alpha_{\tau \mu} \right) 
\biggl\{
s^2_{23} \sin^2 2\phi 
\sin^2 \frac{ ( h_{3} - h_{1} ) x }{2} 
+ 2 \cos 2 \theta_{23} 
\left[ 
c^2_{\phi} \sin^2 \frac{ ( h_{3} - h_{2} ) x }{2}
+ s^2_{\phi} \sin^2 \frac{ ( h_{2} - h_{1} ) x }{2} 
\right] 
\biggr\} 
\nonumber \\
&+& 
\mbox{Im} \left( e^{ i \delta } \alpha_{\tau \mu} \right) 
\biggl\{
c^2_{\phi} \sin ( h_{3} - h_{2} ) x 
- s^2_{\phi} \sin ( h_{2} - h_{1} ) x 
\biggr\} 
\biggr] 
\nonumber \\
&-& 
2 \sin^2 2\theta_{23} 
\left( \alpha_{\mu \mu} + \alpha_{\tau \tau} \right) 
\biggl[ 
c^2_{\phi} s^2_{\phi} 
\sin^2 \frac{ ( h_{3} - h_{1} ) x }{2} 
- \left\{
c^2_{\phi} \sin^2 \frac{ ( h_{3} - h_{2} ) x }{2}
+ s^2_{\phi} \sin^2 \frac{ ( h_{2} - h_{1} ) x }{2} 
\right\} 
\biggr] 
\nonumber \\
&+& 
s_{23} \sin 2\theta_{23} \sin 2\phi 
\biggl[
\mbox{Re} \left( \alpha_{\tau e} \right) 
\left\{
\cos 2 \phi \sin^2 \frac{ ( h_{3} - h_{1} ) x }{2} 
+ \sin^2 \frac{ ( h_{3} - h_{2} ) x }{2}
- \sin^2 \frac{ ( h_{2} - h_{1} ) x }{2} 
\right\} 
\nonumber \\
&-& 
2 \mbox{Im} \left( \alpha_{\tau e} \right) 
\sin \frac{ ( h_{3} - h_{1} ) x }{2} 
\sin \frac{ ( h_{1} - h_{2} ) x }{2} 
\sin \frac{ ( h_{2} - h_{3} ) x }{2} 
\biggr] 
\nonumber \\
&+& 
2 \alpha_{\mu \mu} 
\biggl[ 
2 
+ s^2_{23} \cos 2\theta_{23} \sin^2 2\phi 
\sin^2 \frac{ ( h_{3} - h_{1} ) x }{2} 
- 2 \sin^2 2\theta_{23}
\left\{ 
c^2_{\phi} \sin^2 \frac{ ( h_{3} - h_{2} ) x }{2} 
+ s^2_{\phi} \sin^2 \frac{ ( h_{2} - h_{1} ) x }{2} 
\right\}
\biggr] 
\nonumber \\
&+& 
2 s_{23} \sin 2\phi 
\biggl[
\mbox{Re} \left( e^{- i \delta } \alpha_{\mu e} \right) 
\left\{ 
s^2_{23} \cos 2 \phi 
\sin^2 \frac{ ( h_{3} - h_{1} ) x }{2} 
- c^2_{23} 
\left[ \sin^2 \frac{ ( h_{3} - h_{2} ) x }{2} - \sin^2 \frac{ ( h_{2} - h_{1} ) x }{2} 
\right] 
\right\} 
\nonumber \\
&+& 
\mbox{Im} \left( e^{- i \delta } \alpha_{\mu e} \right) 
\left\{
\cos \frac{ ( h_{3} - h_{1} ) x }{2} 
+ 2 c^2_{23} 
\sin \frac{ ( h_{1} - h_{2} ) x }{2} 
\sin \frac{ ( h_{2} - h_{3} ) x }{2} 
\right\}
\sin \frac{ ( h_{3} - h_{1} ) x }{2} 
\biggr]
\nonumber \\
&+&
2 s^2_{23} \sin^2 2\phi 
\alpha_{e e} 
\sin^2 \frac{ ( h_{3} - h_{1} ) x }{2}, 
\label{xi-1st-order}
\end{eqnarray}
where $h_{i}$ ($i=1,2,3$) denote the eigenvalues of the unperturbed Hamiltonian and $\phi$ is $\theta_{13}$ in matter with $s_{\phi} \equiv \sin \phi$ etc.~\cite{Martinez-Soler:2018lcy,Minakata:2015gra}. 

We restrict ourselves into the order of magnitude estimation of $\xi$.\footnote{
For a better estimation of $\xi$ bound we may need to obtain multi-dimensional $\alpha$ parameter manifold, and information of phases of the $\alpha$ parameters to know if cancellation among the effect of different $\alpha$ parameters occur.  
}
For this purpose we rely on the existing constraints on the $\alpha_{\beta \gamma}$ parameters summarized in Table 2 of Ref.~\cite{Blennow:2016jkn}. We quote here some of the bounds in Table 2. 
$| \alpha_{\mu \mu} | < 2.2 \times 10^{-2}$,  
$| \alpha_{\tau \mu} | < 6.6 \times 10^{-2}$, and  
$| \alpha_{\tau \tau} | < 1.0 \times 10^{-1}$
all for $\Delta m^2 \gsim 0.1$ eV$^2$. 
$| \alpha_{\mu e} | < 3.2 \times 10^{-2}$, and 
$| \alpha_{\tau e} | < 6.9 \times 10^{-2}$ both for $\Delta m^2 \gsim 4$ eV$^2$.
Therefore, it appears that roughly speaking $\xi \lsim 0.1$. 

\section*{Acknowledgements}

We thank Alex Himmel and Mark Messier for very informative correspondences on the NOvA experiments with many helpful suggestions without which we were not able to reach our analysis framework in its current form. We are benefited by useful communications with Pedro Machado, Kimihiro Okumura, and Masashi Yokoyama. Fermilab is operated by the Fermi Research Alliance, LLC under Contract No. DE-AC02-07CH11359 with the United States Department of Energy.  

\bibliographystyle{kpmod}

\begingroup\raggedright\begin{thebibliography}{0}
\expandafter\ifx\csname natexlab\endcsname\relax\def\natexlab#1{#1}\fi

\end{thebibliography}\endgroup


\begin{thebibliography}{99}
\bibitem{Zyla:2020zbs}
P.~A.~Zyla \textit{et al.} [Particle Data Group],
``Review of Particle Physics,''
PTEP \textbf{2020} (2020) no.8, 083C01
doi:10.1093/ptep/ptaa104

\bibitem{Defranchis:2021eos}
M.~M.~Defranchis [ATLAS and CMS],
``Top quark mass and cross sections in ATLAS and CMS,''
[arXiv:2105.05776 [hep-ex]].

\bibitem{CDF:1995wbb}
F.~Abe \textit{et al.} [CDF],
``Observation of top quark production in $\bar{p}p$ collisions,''
Phys. Rev. Lett. \textbf{74} (1995), 2626-2631
doi:10.1103/PhysRevLett.74.2626
[arXiv:hep-ex/9503002 [hep-ex]].

\bibitem{D0:1995jca}
S.~Abachi \textit{et al.} [D0],
``Observation of the top quark,''
Phys. Rev. Lett. \textbf{74} (1995), 2632-2637
doi:10.1103/PhysRevLett.74.2632
[arXiv:hep-ex/9503003 [hep-ex]].

\bibitem{Nambu:1989jt}
Y.~Nambu, 
``Bootstrap Symmetry Breaking in Electroweak Unification'', EFI-89-08.

\bibitem{Miransky:1989ds}
V.~A.~Miransky, M.~Tanabashi and K.~Yamawaki,
``Is the t Quark Responsible for the Mass of W and Z Bosons?,''
Mod. Phys. Lett. A \textbf{4} (1989), 1043
doi:10.1142/S0217732389001210 

\bibitem{Bardeen:1989ds}
W.~A.~Bardeen, C.~T.~Hill and M.~Lindner,
``Minimal Dynamical Symmetry Breaking of the Standard Model,''
Phys. Rev. D \textbf{41} (1990), 1647
doi:10.1103/PhysRevD.41.1647

\bibitem{CMS:2019rvj}
A.~M.~Sirunyan \textit{et al.} [CMS],
``Search for production of four top quarks in final states with same-sign or multiple leptons in proton-proton collisions at $\sqrt{s}=$ 13 TeV,''
Eur. Phys. J. C \textbf{80} (2020) no.2, 75
doi:10.1140/epjc/s10052-019-7593-7
[arXiv:1908.06463 [hep-ex]].

\bibitem{DONuT:2007bsg}
K.~Kodama \textit{et al.} [DONuT],
``Final tau-neutrino results from the DONuT experiment,''
Phys. Rev. D \textbf{78} (2008), 052002
doi:10.1103/PhysRevD.78.052002
[arXiv:0711.0728 [hep-ex]].

\bibitem{OPERA:2018nar}
N.~Agafonova \textit{et al.} [OPERA],
``Final Results of the OPERA Experiment on $\nu_\tau$ Appearance in the CNGS Neutrino Beam,''
Phys. Rev. Lett. \textbf{120} (2018) no.21, 211801
[erratum: Phys. Rev. Lett. \textbf{121} (2018) no.13, 139901]
doi:10.1103/PhysRevLett.120.211801
[arXiv:1804.04912 [hep-ex]].

\bibitem{Super-Kamiokande:2017edb}
Z.~Li \textit{et al.} [Super-Kamiokande],
``Measurement of the tau neutrino cross section in atmospheric neutrino oscillations with Super-Kamiokande,''
Phys. Rev. D \textbf{98} (2018) no.5, 052006
doi:10.1103/PhysRevD.98.052006
[arXiv:1711.09436 [hep-ex]].

\bibitem{IceCube:2019dqi}
M.~G.~Aartsen \textit{et al.} [IceCube],
``Measurement of Atmospheric Tau Neutrino Appearance with IceCube DeepCore,''
Phys. Rev. D \textbf{99} (2019) no.3, 032007
doi:10.1103/PhysRevD.99.032007
[arXiv:1901.05366 [hep-ex]].

\bibitem{Hyper-Kamiokande:2018ofw}
K.~Abe \textit{et al.} [Hyper-Kamiokande],
``Hyper-Kamiokande Design Report,''
[arXiv:1805.04163 [physics.ins-det]].

\bibitem{DUNE:2020ypp}
B.~Abi \textit{et al.} [DUNE],
``Deep Underground Neutrino Experiment (DUNE), Far Detector Technical Design Report, Volume II: DUNE Physics,''
[arXiv:2002.03005 [hep-ex]].

\bibitem{DeGouvea:2019kea}
A.~De Gouv\^ea, K.~J.~Kelly, G.~V.~Stenico and P.~Pasquini,
``Physics with Beam Tau-Neutrino Appearance at DUNE,''
Phys. Rev. D \textbf{100} (2019) no.1, 016004
doi:10.1103/PhysRevD.100.016004
[arXiv:1904.07265 [hep-ph]].

\bibitem{Machado:2020yxl}
P.~Machado, H.~Schulz and J.~Turner,
``Tau neutrinos at DUNE: New strategies, new opportunities,''
Phys. Rev. D \textbf{102} (2020) no.5, 053010
doi:10.1103/PhysRevD.102.053010
[arXiv:2007.00015 [hep-ph]].

\bibitem{Ghoshal:2019pab}
A.~Ghoshal, A.~Giarnetti and D.~Meloni,
``On the role of the $\nu_{\tau}$ appearance in DUNE in constraining standard neutrino physics and beyond,''
JHEP \textbf{12} (2019), 126
doi:10.1007/JHEP12(2019)126
[arXiv:1906.06212 [hep-ph]]. 

\bibitem{T2K:2019bcf}
K.~Abe \textit{et al.} [T2K],
``Constraint on the matter\textendash{}antimatter symmetry-violating phase in neutrino oscillations,''
Nature \textbf{580} (2020) no.7803, 339-344
[erratum: Nature \textbf{583} (2020) no.7814, E16]
doi:10.1038/s41586-020-2177-0
[arXiv:1910.03887 [hep-ex]].

\bibitem{NOvA:2019cyt}
M.~A.~Acero \textit{et al.} [NOvA],
``First Measurement of Neutrino Oscillation Parameters using Neutrinos and Antineutrinos by NOvA,''
Phys. Rev. Lett. \textbf{123} (2019) no.15, 151803
doi:10.1103/PhysRevLett.123.151803
[arXiv:1906.04907 [hep-ex]].

\bibitem{Martinez-Soler:2018lcy}
I.~Martinez-Soler and H.~Minakata,
``Standard versus Non-Standard CP Phases in Neutrino Oscillation in Matter with Non-Unitarity,''
PTEP \textbf{2020} (2020) no.6, 063B01
doi:10.1093/ptep/ptaa062
[arXiv:1806.10152 [hep-ph]].


\bibitem{Wolfenstein:1977ue}
L.~Wolfenstein,
``Neutrino Oscillations in Matter,''
Phys. Rev. D \textbf{17} (1978), 2369-2374
doi:10.1103/PhysRevD.17.2369

\bibitem{Ohlsson:2012kf}
T.~Ohlsson,
``Status of non-standard neutrino interactions,''
Rept. Prog. Phys. \textbf{76} (2013), 044201
doi:10.1088/0034-4885/76/4/044201
[arXiv:1209.2710 [hep-ph]].

\bibitem{Miranda:2015dra}
O.~G.~Miranda and H.~Nunokawa,
``Non standard neutrino interactions: current status and future prospects,''
New J. Phys. \textbf{17} (2015) no.9, 095002
doi:10.1088/1367-2630/17/9/095002
[arXiv:1505.06254 [hep-ph]].

\bibitem{Farzan:2017xzy}
Y.~Farzan and M.~Tortola,
``Neutrino oscillations and Non-Standard Interactions,''
Front. in Phys. \textbf{6} (2018), 10
doi:10.3389/fphy.2018.00010
[arXiv:1710.09360 [hep-ph]].

\bibitem{Antusch:2008tz}
S.~Antusch, J.~P.~Baumann and E.~Fernandez-Martinez,
``Non-Standard Neutrino Interactions with Matter from Physics Beyond the Standard Model,''
Nucl. Phys. B \textbf{810} (2009), 369-388
doi:10.1016/j.nuclphysb.2008.11.018
[arXiv:0807.1003 [hep-ph]].

\bibitem{Biggio:2009nt}
C.~Biggio, M.~Blennow and E.~Fernandez-Martinez,
``General bounds on non-standard neutrino interactions,''
JHEP \textbf{08} (2009), 090
doi:10.1088/1126-6708/2009/08/090
[arXiv:0907.0097 [hep-ph]].

\bibitem{Esteban:2018ppq}
I.~Esteban, M.~C.~Gonzalez-Garcia, M.~Maltoni, I.~Martinez-Soler and J.~Salvado,
``Updated constraints on non-standard interactions from global analysis of oscillation data,''
JHEP \textbf{08} (2018), 180
doi:10.1007/JHEP08(2018)180
[arXiv:1805.04530 [hep-ph]].


\bibitem{LSND:2001aii}
A.~Aguilar-Arevalo \textit{et al.} [LSND],
``Evidence for neutrino oscillations from the observation of $\bar{\nu}_e$ appearance in a $\bar{\nu}_\mu$
 beam,''
Phys. Rev. D \textbf{64} (2001), 112007
doi:10.1103/PhysRevD.64.112007
[arXiv:hep-ex/0104049 [hep-ex]].

\bibitem{MiniBooNE:2013uba}
A.~A.~Aguilar-Arevalo \textit{et al.} [MiniBooNE],
``Improved Search for $\bar \nu_\mu \rightarrow \bar \nu_e$ Oscillations in the MiniBooNE Experiment,''
Phys. Rev. Lett. \textbf{110} (2013), 161801
doi:10.1103/PhysRevLett.110.161801
[arXiv:1303.2588 [hep-ex]].

\bibitem{Dasgupta:2021ies}
B.~Dasgupta and J.~Kopp,
``Sterile Neutrinos,''
Phys. Rept. \textbf{928} (2021), 1-63
doi:10.1016/j.physrep.2021.06.002
[arXiv:2106.05913 [hep-ph]].

\bibitem{Dentler:2018sju}
M.~Dentler, \'A.~Hern\'andez-Cabezudo, J.~Kopp, P.~A.~N.~Machado, M.~Maltoni, I.~Martinez-Soler and T.~Schwetz,
``Updated Global Analysis of Neutrino Oscillations in the Presence of eV-Scale Sterile Neutrinos,''
JHEP \textbf{08} (2018), 010
doi:10.1007/JHEP08(2018)010
[arXiv:1803.10661 [hep-ph]]. 

\bibitem{Conrad:2013mka}
J.~M.~Conrad, W.~C.~Louis and M.~H.~Shaevitz,
``The LSND and MiniBooNE Oscillation Searches at High $\Delta m^2$,''
Ann. Rev. Nucl. Part. Sci. \textbf{63} (2013), 45-67
doi:10.1146/annurev-nucl-102711-094957
[arXiv:1306.6494 [hep-ex]].

\bibitem{Machado:2019oxb}
P.~A.~Machado, O.~Palamara and D.~W.~Schmitz,
``The Short-Baseline Neutrino Program at Fermilab,''
Ann. Rev. Nucl. Part. Sci. \textbf{69} (2019), 363-387
doi:10.1146/annurev-nucl-101917-020949
[arXiv:1903.04608 [hep-ex]].

\bibitem{JSNS2:2021hyk}
S.~Ajimura \textit{et al.} [JSNS2],
``The JSNS2 detector,''
Nucl. Instrum. Meth. A \textbf{1014} (2021), 165742
doi:10.1016/j.nima.2021.165742
[arXiv:2104.13169 [physics.ins-det]]. 


\bibitem{Minkowski:1977sc}
P.~Minkowski,
``$\mu \to e\gamma$ at a Rate of One Out of $10^{9}$ Muon Decays?,''
Phys. Lett. B \textbf{67} (1977), 421-428
doi:10.1016/0370-2693(77)90435-X

\bibitem{Yanagida:1979as}
T.~Yanagida,
``Horizontal gauge symmetry and masses of neutrinos,''
Conf. Proc. C \textbf{7902131} (1979), 95-99
KEK-79-18-95; 
T. Yanagida, {\it Workshop on the Baryon Number of the Universe and Unified Theories, Tsukuba, Japan, 1979, (C79-02-13.1)} (KEK library, 1979).

\bibitem{Gell-Mann:1979vob}
M.~Gell-Mann, P.~Ramond and R.~Slansky,
``Complex Spinors and Unified Theories,''
Conf. Proc. C \textbf{790927} (1979), 315-321
[arXiv:1306.4669 [hep-th]].

\bibitem{Glashow:1979nm}
S.~L.~Glashow,
``The Future of Elementary Particle Physics,''
NATO Sci. Ser. B \textbf{61} (1980), 687
doi:10.1007/978-1-4684-7197-7\_15

\bibitem{Mohapatra:1979ia}
R.~N.~Mohapatra and G.~Senjanovic,
``Neutrino Mass and Spontaneous Parity Nonconservation,''
Phys. Rev. Lett. \textbf{44} (1980), 912
doi:10.1103/PhysRevLett.44.912


\bibitem{Antusch:2006vwa}
S.~Antusch, C.~Biggio, E.~Fernandez-Martinez, M.~B.~Gavela and J.~Lopez-Pavon,
``Unitarity of the Leptonic Mixing Matrix,''
JHEP \textbf{10} (2006), 084
doi:10.1088/1126-6708/2006/10/084
[arXiv:hep-ph/0607020 [hep-ph]].

\bibitem{Fong:2016yyh}
C.~S.~Fong, H.~Minakata and H.~Nunokawa,
``A framework for testing leptonic unitarity by neutrino oscillation experiments,''
JHEP \textbf{02} (2017), 114
doi:10.1007/JHEP02(2017)114
[arXiv:1609.08623 [hep-ph]].

\bibitem{Antusch:2014woa}
S.~Antusch and O.~Fischer,
``Non-unitarity of the leptonic mixing matrix: Present bounds and future sensitivities,''
JHEP \textbf{10} (2014), 094
doi:10.1007/JHEP10(2014)094
[arXiv:1407.6607 [hep-ph]].

\bibitem{Escrihuela:2015wra}
F.~J.~Escrihuela, D.~V.~Forero, O.~G.~Miranda, M.~Tortola and J.~W.~F.~Valle,
``On the description of nonunitary neutrino mixing,''
Phys. Rev. D \textbf{92} (2015) no.5, 053009
[erratum: Phys. Rev. D \textbf{93} (2016) no.11, 119905]
doi:10.1103/PhysRevD.92.053009
[arXiv:1503.08879 [hep-ph]].

\bibitem{Fernandez-Martinez:2016lgt}
E.~Fernandez-Martinez, J.~Hernandez-Garcia and J.~Lopez-Pavon,
``Global constraints on heavy neutrino mixing,''
JHEP \textbf{08} (2016), 033
doi:10.1007/JHEP08(2016)033
[arXiv:1605.08774 [hep-ph]].

\bibitem{Blennow:2016jkn}
M.~Blennow, P.~Coloma, E.~Fernandez-Martinez, J.~Hernandez-Garcia and J.~Lopez-Pavon,
``Non-Unitarity, sterile neutrinos, and Non-Standard neutrino Interactions,''
JHEP \textbf{04} (2017), 153
doi:10.1007/JHEP04(2017)153
[arXiv:1609.08637 [hep-ph]]. 

\bibitem{Schechter:1980gr}
J.~Schechter and J.~W.~F.~Valle,
``Neutrino Masses in SU(2) x U(1) Theories,''
Phys. Rev. D \textbf{22} (1980), 2227
doi:10.1103/PhysRevD.22.2227

\bibitem{Barger:1980tfa}
V.~D.~Barger, P.~Langacker, J.~P.~Leveille and S.~Pakvasa,
``Consequences of Majorana and Dirac Mass Mixing for Neutrino Oscillations,''
Phys. Rev. Lett. \textbf{45} (1980), 692
doi:10.1103/PhysRevLett.45.692

\bibitem{Fong:2017gke}
C.~S.~Fong, H.~Minakata and H.~Nunokawa,
``Non-unitary evolution of neutrinos in matter and the leptonic unitarity test,''
JHEP \textbf{02} (2019), 015
doi:10.1007/JHEP02(2019)015
[arXiv:1712.02798 [hep-ph]]. 

\bibitem{Parke:2015goa}
S.~Parke and M.~Ross-Lonergan,
``Unitarity and the three flavor neutrino mixing matrix,''
Phys. Rev. D \textbf{93} (2016) no.11, 113009
doi:10.1103/PhysRevD.93.113009
[arXiv:1508.05095 [hep-ph]].

\bibitem{Ellis:2020hus}
S.~A.~R.~Ellis, K.~J.~Kelly and S.~W.~Li,
``Current and Future Neutrino Oscillation Constraints on Leptonic Unitarity,''
JHEP \textbf{12} (2020), 068
doi:10.1007/JHEP12(2020)068
[arXiv:2008.01088 [hep-ph]].


\bibitem{NOvA:2018gge}
M.~A.~Acero \textit{et al.} [NOvA],
``New constraints on oscillation parameters from $\nu_e$ appearance and $\nu_\mu$ disappearance in the NOvA experiment,''
Phys. Rev. D \textbf{98} (2018), 032012
doi:10.1103/PhysRevD.98.032012
[arXiv:1806.00096 [hep-ex]].

\bibitem{Vinton:2018aqq}
L.~Vinton,
``Measurement of Muon Neutrino Disappearance with the NOvA Experiment,''
doi:10.2172/1423216.

\bibitem{Himmel-Nu2020}
A.~Himmel, Plenary Talk at Neutrino 2020, 
``New Oscillation Results from the NOvA Experiment'', 
Zenodo. http://doi.org/10.5281/zenodo.4142045.

\bibitem{NOvA:2021nfi}
M.~A.~Acero \textit{et al.} [NOvA and R. Group],
``An Improved Measurement of Neutrino Oscillation Parameters by the NOvA Experiment,''
[arXiv:2108.08219 [hep-ex]].

\bibitem{NOvA-data} 
https://novaexperiment.fnal.gov/data-releases/

\bibitem{Super-Kamiokande:2017yvm}
K.~Abe \textit{et al.} [Super-Kamiokande],
``Atmospheric neutrino oscillation analysis with external constraints in Super-Kamiokande I-IV,''
Phys. Rev. D \textbf{97} (2018) no.7, 072001
doi:10.1103/PhysRevD.97.072001
[arXiv:1710.09126 [hep-ex]].

\bibitem{Stuttard:2020zsj}
T.~Stuttard [IceCube],
``Neutrino oscillations and PMNS unitarity with IceCube/DeepCore and the IceCube Upgrade,''
PoS \textbf{NuFact2019} (2020), 099
doi:10.22323/1.369.0099

\bibitem{IceCubeCollaboration:2021euf}
R.~Abbasi \textit{et al.} [(IceCube Collaboration)* and IceCube],
``All-flavor constraints on nonstandard neutrino interactions and generalized matter potential with three years of IceCube DeepCore data,''
Phys. Rev. D \textbf{104} (2021) no.7, 072006
doi:10.1103/PhysRevD.104.072006
[arXiv:2106.07755 [hep-ex]].

\bibitem{KM3NeT:2021ozk}
S.~Aiello \textit{et al.} [KM3NeT],
``Determining the Neutrino Mass Ordering and Oscillation Parameters with KM3NeT/ORCA,''
[arXiv:2103.09885 [hep-ex]].

\bibitem{JUNO:2015zny}
F.~An \textit{et al.} [JUNO],
``Neutrino Physics with JUNO,''
J. Phys. G \textbf{43} (2016) no.3, 030401
doi:10.1088/0954-3899/43/3/030401
[arXiv:1507.05613 [physics.ins-det]].

\bibitem{T2K:2012bge}
K.~Abe \textit{et al.} [T2K],
``T2K neutrino flux prediction,''
Phys. Rev. D \textbf{87} (2013) no.1, 012001
doi:10.1103/PhysRevD.87.012001
[arXiv:1211.0469 [hep-ex]]. 

\bibitem{MINERvA:2016iqn}
L.~Aliaga \textit{et al.} [MINERvA],
``Neutrino Flux Predictions for the NuMI Beam,''
Phys. Rev. D \textbf{94} (2016) no.9, 092005
doi:10.1103/PhysRevD.94.092005
[arXiv:1607.00704 [hep-ex]].

\bibitem{Marshall:2019vdy}
C.~M.~Marshall, K.~S.~McFarland and C.~Wilkinson,
``Neutrino-electron elastic scattering for flux determination at the DUNE oscillation experiment,''
Phys. Rev. D \textbf{101} (2020) no.3, 032002
doi:10.1103/PhysRevD.101.032002
[arXiv:1910.10996 [hep-ex]].

\bibitem{MINERvA:2019hhc}
E.~Valencia \textit{et al.} [MINERvA],
``Constraint of the MINER$\nu$A medium energy neutrino flux using neutrino-electron elastic scattering,''
Phys. Rev. D \textbf{100} (2019) no.9, 092001
doi:10.1103/PhysRevD.100.092001
[arXiv:1906.00111 [hep-ex]].

\bibitem{Minakata:2015gra}
H.~Minakata and S.~J.~Parke,
``Simple and Compact Expressions for Neutrino Oscillation Probabilities in Matter,''
JHEP \textbf{01} (2016), 180
doi:10.1007/JHEP01(2016)180
[arXiv:1505.01826 [hep-ph]]. 

\end{thebibliography}

\end{document}